# Strain-tunable anomalous Hall plateau in antiferromagnet $CoNb_3S_6$

Long Chen[1], Richard Lai[1], Shashi Pandey[1], Dapeng Cui[1], Alexander Brassington[1], Jian Liu[1, *], Haidong Zhou[1, *]

[1]Department of Physics and Astronomy, University of Tennessee, Knoxville, TN, USA

*Corresponding author. Email: jianliu@utk.edu; hzhou10@utk.edu;

**Abstract:**

Antiferromagnets exhibiting the anomalous Hall effect represent a fascinating convergence of magnetism, topology, and electronic structure. Identifying antiferromagnets with large and tunable anomalous Hall effects is crucial for the development of spintronic applications. Here, we report a strain-tunable anomalous Hall plateau in $CoNb_3S_6$, which is a prime candidate for altermagnetism. The plateau emerges as a flat extended intermediate step of the anomalous Hall hysteresis loop with a controllable step height with temperature and strain. The remarkable tunability of the plateau position is in contrast with typical magnetic plateau associated with a field-induced metastable magnetic structure, but indicates the existence of a hidden phase transition that significantly alters the magnetic anisotropy energy without changing the magnetic order. The symmetry analysis of the strain tuning suggests that the hidden phase preserves the rotational symmetry of the *ab*-plane. Our results show the plateau reflects the phase coexistence during the hidden transition, and anomalous Hall resistivity of the plateau is thus non-volatile, enabling a novel four-state switching of the anomalous Hall effect.

**Introduction**

Anomalous Hall effect (AHE), which appears as a magnetic hysteresis loop of the Hall resistivity, is a result of the interplay between magnetic ordering, spin-orbit coupling, and charge transport in materials [1]. While traditionally known as a property of ferromagnets [2-5], AHE has been observed in an increasing number of antiferromagnets [6-14], defying the empirical expectation of significant magnetization and facilitating the discovery of new topological phases [15-17]. The fact that the AHE resistivity switches sign in an antiferromagnet captures the reversal of the Néel order, which makes this phenomenon particularly useful for antiferromagnetic spintronics, because it allows an effective detection of the Néel vector despite the absence of net magnetization [18,19]. Antiferromagnets with large and controllable AHE is thus highly desirable for spintronic applications that aim to exploit the ultra-fast spin dynamics and vanishing stray field of antiferromagnets [19-21].

$CoNb_3S_6$ is an excellent example of layered antiferromagnets [22,23] that has attracted increasing attentions in recent years due to its rich physics [24-29]. Nirmal et al. [30] first discovered the unexpected large AHE (~27 $\Omega^{-1}$ cm$^{-1}$) and later Giulia et al. [31] found the AHE can be further enhanced in small exfoliated crystals (~400 $\Omega^{-1}$ cm$^{-1}$). In particular, the Néel order is almost fully compensated, with a tiny magnetization that is too small to account for the large AHE by conventional ferromagnetism [30]. In 2020, Smejkal et al. [32] rebranded the AHE in $CoNb_3S_6$ as the "crystal Hall effect", which may occur in collinear antiferromagnets when nonmagnetic atoms occupy non-centrosymmetric positions around magnetic sites and form an alternating lattice unit. This class of antiferromagnets was later dubbed "altermagnets" [33,34], with $CoNb_3S_6$ being an early candidate, along with several other materials [35-37]. Meanwhile, neutron scattering measurements suggested that $CoNb_3S_6$ has a non-coplanar antiferromagnetic structure and the AHE can be attributed to spin chirality [38], i.e., a spontaneous topological Hall effect. In either case, the switching of the AHE resistivity signifies the reversal of the underlying magnetic order parameter, which is experimentally achieved by applying a magnetic field to flip the small canted moment. The coercivity field thus rapidly increases with decreasing temperature. However, the increase in coercivity appears to be nonuniform, and the AHE hysteresis loop from multiple reports [24,28,30,31,39] shows a jump-like feature, which is typical of metamagnetic transitions [40,41]. Since metamagnetism is a signature of field-induced magnetic structure, the nature of this behavior has significant implications to the magnetism of this material and the exact process of its magnetic reversal, but has yet to be investigated in detail.

In this study, we show the AHE coercivity of $CoNb_3S_6$ increases through a big jump, resulting in an anomalous Hall plateau that we found to be highly tunable. Our analysis with different strain configurations indicates that the coercivity jump is due to a hidden phase transition, during which the anisotropy energy is significantly enhanced without breaking or changing rotational symmetry. Since the transition temperature is sensitive to mechanical stress and both phases with large and small coercivity coexist during the transition, the plateau resistivity can be tuned to any value between the positive and negative polarities of the AHE, and it is non-volatile. The magnetic reversal thus exhibits an unusual four-state switching process that switches the AHE between the two polarities through two variable plateaus.

## Results and Discussion

### 1. Anomalous Hall plateau and strain tuning.

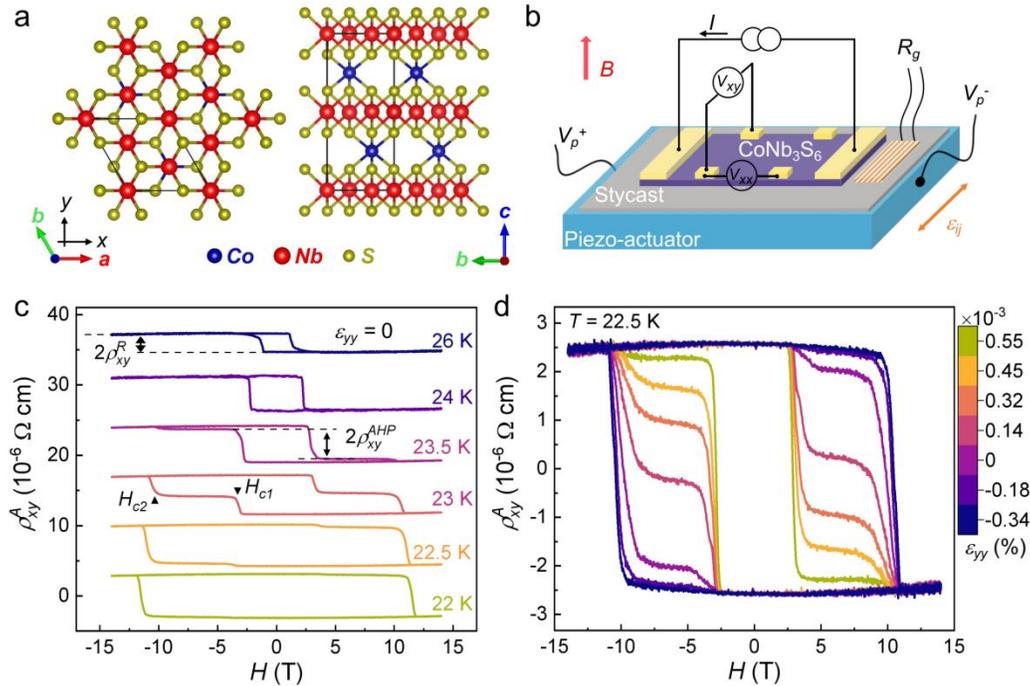

**Figure 1. Strain tunable anomalous Hall plateau.** (a) Top (left panel) and side (right panel) views of the crystal structure of $CoNb_3S_6$. (b) Schematic illustration of the strained $CoNb_3S_6$

device and corresponding configuration of Hall measurement ($\rho_{xy} = V_{xy}/I$). $\varepsilon_{xx}$ and $\varepsilon_{yy}$ denote the uniaxial strain along and perpendicular to the *a*-axis, respectively. The magnetic field is always perpendicular to the *ab* plane. (c) Anomalous Hall resistivity ($\rho^{A}_{xy}$) up to 14 T at different temperatures after the sample (S#1) is mounted on the piezo-actuator without applying voltage ($V_p = 0$ V, $\varepsilon_{yy} = 0$). (d) Anomalous Hall resistivity up to 14 T at $T = 22.5$ K with strain varying from $-3.4 \times 10^{-3}$ % to $5.5 \times 10^{-3}$ %. Strain is applied perpendicular to the *a*-axis ($\varepsilon_{yy}$).

Fig. 1 illustrates the experimental setup and the key observations. Owing to its layered structure (Fig. 1a), CoNb$_3$S$_6$ single crystals were cleaved into thin plates (thickness ~ 50 μm) and attached to a piezo-actuator by Stycast (Fig. 1b). Upon increasing the voltage, the piezo-actuator expands to apply a tensile uniaxial strain to the sample, which is monitored by a strain gauge attached to the piezo-actuator [42,43]. Even at zero voltage, the AHE already shows an interesting behavior below the Néel transition $T_N \sim 28$ K as one can clearly see the development of a double-loop-like hysteresis with two coercivity fields ($H_{c1}$ and $H_{c2}$) from the originally single loop as temperature decreases (Fig. 1c). The anomalous Hall resistivity sharply jumps at both $H_{c1}$ and $H_{c2}$ with a flat intermediate step in between. The primary effect of decreasing the temperature is to change the height of the intermediate step from one polarity of the AHE to the other. In other words, the AHE resistivity jump of $H_{c2}$ increases at the expense of the jump of $H_{c1}$. As the jump vanishes at $H_{c1}$, the hysteresis recovers the single-loop shape but with a much larger coercivity at $H_{c2}$. Therefore, the increase of coercivity is discontinuous with a ~6 T gap between $H_{c1}$ and $H_{c2}$. We verified this double-loop-like hysteresis on a free-standing sample in both AHE and magnetization, indicating that this behavior is associated with the reversal of the underlying Néel vector (See Supplementary Fig. S1-S7, Table S1-S2). Such a double-loop-like hysteresis can be in fact spotted by taking a close inspection of the previously reported AHE data [24,28,30,31,39]. The similar behavior in AHE resistivity and other physical properties among several reports (Table S3) confirms our observation of the double-loop-like hysteresis, but its systematic evolution has not been resolved.

What is even more interesting is that the intermediate step height can be tuned by the uniaxial strain. In the measurement shown in Fig. 1d, the uniaxial strain ($\varepsilon_{yy}$) was perpendicular to the *a*-axis. When cooling the sample to the same temperature ($T = 22.5$ K) with different voltages applied to the piezo-actuator ($-3.4 \times 10^{-3}$ % $\leq \varepsilon_{yy} \leq 5.5 \times 10^{-3}$ %), the step position systematically shifts from one polarity of the AHE all the way to the other. As a result, the hysteresis loop evolves from a single loop with larger coercivity, to a double loop, and back to a single loop but with smaller coercivity. Throughout this process, $H_{c1}$ and $H_{c2}$ remain largely unchanged, and the saturated AHE resistivity is unchanged as well. The only change is the height of the intermediate step, and the step remains rather flat and is thus dubbed here the *anomalous Hall plateau* (AHP). Magnetic plateau often occurs due to metamagnetic transition in materials where the reversal of the magnetic moments takes more than one step [40,44-46]. When an intermediate step corresponds to a discrete change of the magnetic structure, it often manifests as a plateau. However, that also means the position of the plateau should be fixed [41,47-49], in contrast to our observations. The fact that the observed AHP is continuously variable with temperature and strain implies a different origin.

## 2. Anomalous Hall plateau and hidden phase transition

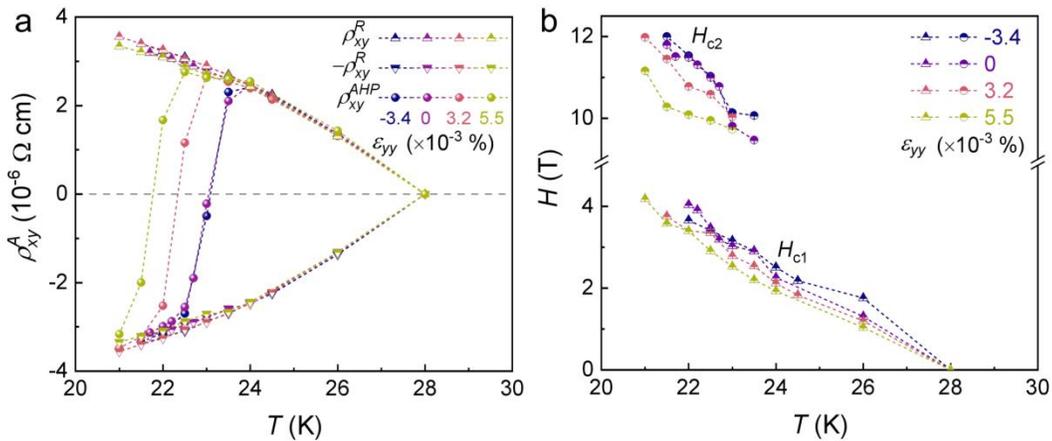

**Figure 2. Potential hidden phase transition and its evolution with strain.** (a) Temperature dependence of remnant anomalous Hall resistivity ($\rho^R_{xy}$) and AHP resistivity ($\rho^{AHP}_{xy}$), (b) critical magnetic fields ($H_{c1}$ and $H_{c2}$) for CoNb$_3$S$_6$ with different strain. Strain is applied perpendicular to the *a*-axis ($\varepsilon_{yy}$).

To reveal the nature of the AHP, we systematically measured the temperature dependence at various strains (Fig. S8) and identified a phase transition at $T^* \sim 23$ K. As shown in Fig. 2a, the remnant Hall resistivity ($\rho^R_{xy} = [\rho^A_{xy} (0\ \text{T}, H\uparrow) - \rho^A_{xy} (0\ \text{T}, H\downarrow)]/2$) emerges around $T_N$ and gradually increases with decreasing temperature as one normally expects from an enlarging order parameter. Even when strain is applied, one can see that $\rho^R_{xy}$ hardly changes, indicating no change to the magnetic structure. In contrast, the AHP exhibits not only a rather different temperature dependence but also significant strain dependence. To illustrate this, we extracted the AHP resistivity as $\rho^{AHP}_{xy} = [\rho^A_{xy}(-7\,T, H\downarrow) - \rho^A_{xy}(+7\,T, H\uparrow)]/2$, where $\rho^A_{xy}(-7\,T, H\downarrow)$ and $\rho^A_{xy}(+7\,T, H\uparrow)$ represents the AHE resistivity on the plateau after saturating the sample at +14 T and -14 T, respectively. One can see $\rho^{AHP}_{xy}$ is initially the same as $\rho^R_{xy}$, but rapidly deviates toward $-\rho^R_{xy}$ as temperature decreases (Fig. 2a). The deviation starts below 24 K when the applied strain is zero, and its full temperature window is ~2 K, which can be better seen as the coexistence temperature range of $H_{c1}$ and $H_{c2}$ (Fig. 2b). This behavior corresponds to the movement of the AHP from one polarity to the other, as seen in Fig. 1d. More importantly, this temperature window shifts to lower temperatures as strain increases while its size remains almost the same. These observations indicate that a phase transition occurs within this temperature window, during which the relative proportions of the high-temperature phase (with smaller coercivity) and the low-temperature phase (with larger coercivity) gradually shift with temperature. Such coexistence suggests characteristics of a first-order transition. Furthermore, this transition is more sensitive to mechanical strain than the AFM transition.

## 3. Anisotropic strain effect and symmetry analysis

What is particularly interesting is that no obvious feature or anomaly was observed in other temperature-dependent thermodynamic properties around $T^*$, including electrical resistivity ($\rho_{xx}$ vs $T$, Fig. S7), magnetic susceptibility ($\chi$ vs $T$, Fig. S3), remnant Hall resistivity, and remnant c-axis magnetization ($\rho^R_{xy}(0\,T)$ & $\Delta M(0\,T)$ vs $T$, Fig. S9). Therefore, this phase transition appears to be *hidden* in most temperature-dependent electrical and magnetic measurements under the static magnetic field. Similarly, no change was observed in the crystal structure around $T^*$ within the resolution of our measurement (Fig. S10). The only indicator other than the AHP is a very weak kink in the temperature dependence of the heat capacity (Fig. S11), confirming the existence of a phase transition. Given that the main change across the transition is a sharp jump in coercivity, this hidden phase below $T^*$ is likely associated with a multipolar order that significantly enhances the magnetic anisotropy energy without changing the magnetic structure.

Such an order often manifests in heat capacity but is invisible in other thermodynamic measurements [50].

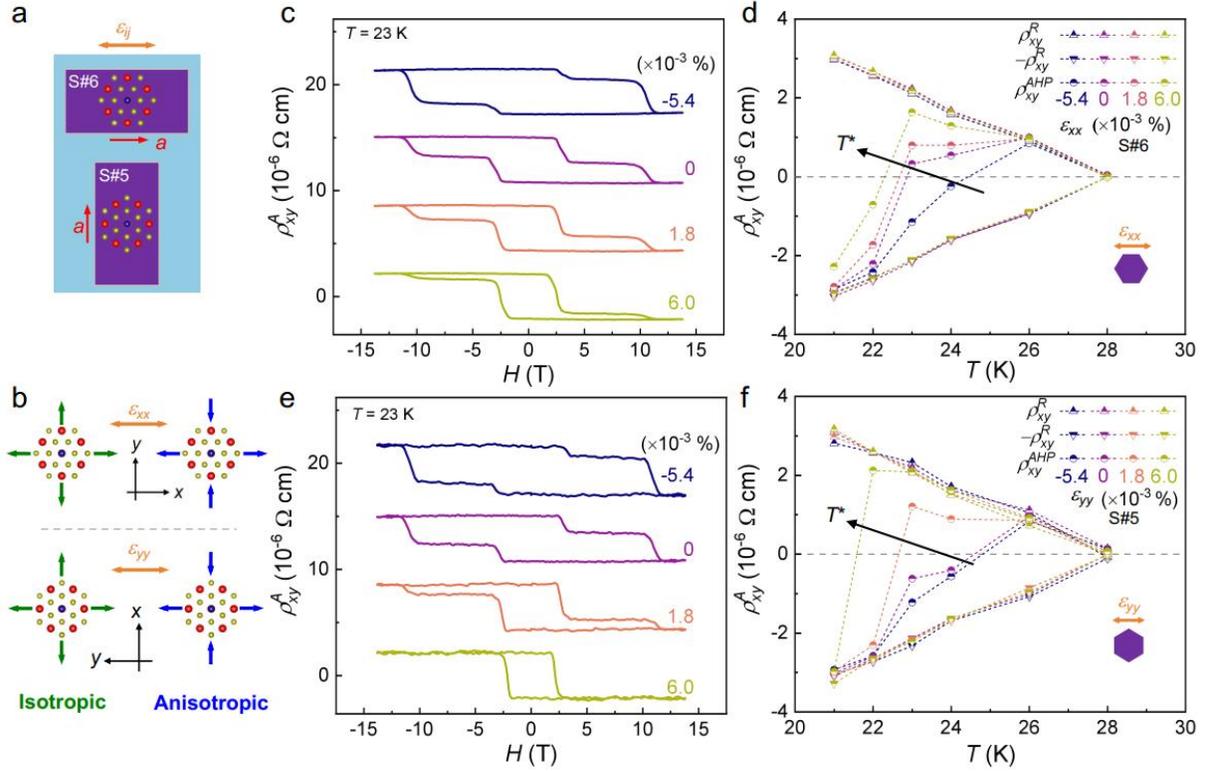

**Figure 3. Anisotropic strain and symmetry analysis.** (a) Schematic illustration of the anisotropic strain applied on two $CoNb_3S_6$ thin plates (S5# and S#6). (b) Schematic isotropic and anisotropic deformations under strain along ($\varepsilon_{xx}$) or perpendicular ($\varepsilon_{yy}$) to the *a*-axis. The arrows represent the extension of strain (orange) and corresponding isotropic (green) and anisotropic (blue) deformations. Anomalous Hall resistivity at *T* = 23 K with different strain (c) along (S#6, $\varepsilon_{xx}$) and (e) perpendicular (S#5, $\varepsilon_{yy}$) to the *a*-axis. Temperature dependence of remnant anomalous Hall resistivity ($\rho^R_{xy}$) and AHP resistivity ($\rho^{AHP}_{xy}$) for $CoNb_3S_6$ with different strain (d) along and (f) perpendicular to the *a*-axis.

To further investigate the symmetry of the hidden phase, we compared the AHP response to uniaxial strain along and perpendicular to the *a*-axis. Typically, uniaxial strain causes two types of deformation at the same time, an isotropic deformation and an anisotropic deformation. The former stands for a uniform expansion or compression of the *ab*-plane, and its effect should not depend on the orientation of the uniaxial strain. The latter, in contrast, breaks the rotational
7

symmetry and effectively lowers the point group symmetry from $D_6$ to $D_2$. Since it changes sign when comparing uniaxial strain parallel and perpendicular to the *a*-axis (Fig. 3a,b), its effect on the AHP should reverse between these two uniaxial strain configurations. In order to minimize extrinsic effects on the comparison, such as thermal strain, variation in piezo-actuators, and sample inhomogeneity, two thin plates (S#5 and S#6) were cleaved from the same parent crystal and attached to the same piezo-actuator simultaneously (Fig. S12-S13) with a 90-degree rotation from each other. As the uniaxial strain increases, one can see that the AHP shifts in the same way for both samples (Fig. 3c, e), demonstrating that the strain tuning is insensitive to the orientation of the uniaxial strain within the *ab*-plane and driven by the isotropic deformation. The temperature dependencies of $\rho^{AHP}_{xy}$ (Fig. 3d, f) and the magnetic critical fields (Fig. S14) both show a downshift of the hidden phase transition for both samples, which is also consistent with the behavior of the sample in Fig. 1. These results indicate that the hidden phase preserves the rotational symmetry, which is in line with the absence of any apparent change to the magnetic order.

4. **Four-state switching**

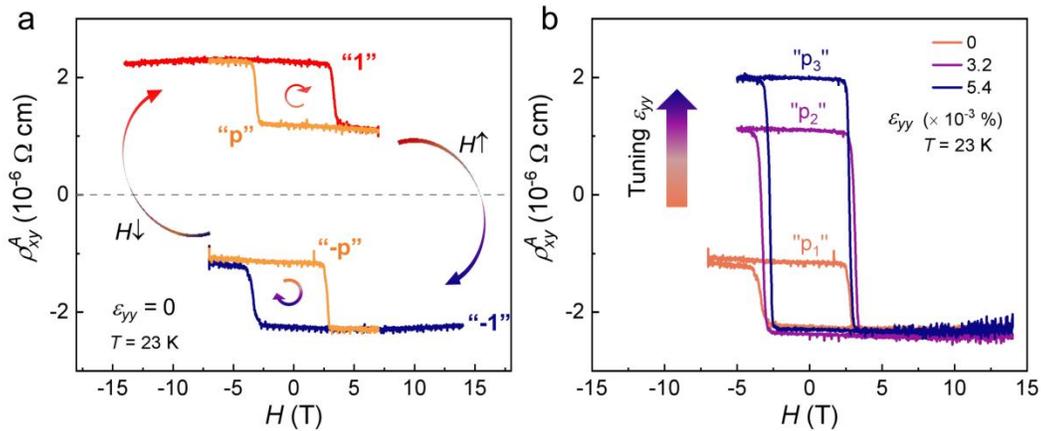

**Figure 4. Potential in achieving muti-value storage.** (a) Schematic diagram of multi-value storage based on four stable states. The arrows show the switching between different states ("±1" and "±p" ) under $\varepsilon_{yy} = 0$. (b) Different AHP states ("p$_1$", "p$_2$" and "p$_3$") under different strain. The strain is changed at 40 K and then the system is cooled down to 23 K.

Having established the hidden transition, the tunable AHP can be understood as a result of the phase coexistence during the transition. In other words, only the high-temperature phase reverses the Néel vector at $H_{c1}$ and the low-temperature phase does not flip until reaching $H_{c2}$, creating the AHP in between. The position of the AHP thus characterizes the relative population of the two phases. And the AHP resistivity should be non-volatile, which is indeed the case as shown in Fig. 4a, where the AHE resistivity remains the same as the AHP resistivity when the field decreases back to zero after only surpassing $H_{c1}$. The non-volatility of the AHP further rules out any metamagnetic transition because an intermediate step reached by a metamagnetic transition is only stable at finite fields [40,44-49].

The non-volatility of the AHP also implies that the AHE resistivity can be switched between the remnant value and the AHP value through a minor loop as shown in Fig. 4a. This behavior effectively enables a four-state switching of the AHE, where the positive and negative remnant value can switch with their corresponding AHP value, respectively, and the switching between these two pairs can be achieved by a full reversal of the sample as illustrated in Fig. 4a. The fact that the AHP value is tunable allows the minor loop switching to occur with a variable state as shown in Fig. 4b. If one encoded the remnant value as "±1", the variable AHP value can be encoded as "±p" that floats between "1" and "-1". Although the ordering temperature of $CoNb_3S_6$ is too low for practical applications, such a four-state AHE switching could inspire novel switching mechanisms and antiferromagnetic devices that integrate both digital and analog signals.

**Conclusions**

In summary, a strain-tunable AHP has been discovered in layered antiferromagnet, $CoNb_3S_6$. The AHP is found to be correlated with the transition of a previously unreported hidden phase, the temperature of which is sensitive to isotropic deformation of the *ab*-plane. Such a plateau is thus distinct from that of a metamagnetic transition and exhibits non-volatility, which opens new opportunities to achieve novel switching and multi-value storage. Further investigations into the underlying mechanisms (like resonant X-ray scattering, high-angle neutron scattering, ultrasonic or optical measurements) and potential applications of this strain-engineered magnetic phenomenon could contribute to the advancement and innovations in data storage and other magnetic technologies.

**Methods**

**Single Crystal Growth, Structure Characterization and Composition Analysis.** $CoNb_3S_6$ single crystals were grown by a chemical vapor transportation method using $I_2$ as transport agency [22]. Hexagonal-shaped, shiny-silver single crystals with sizes up to 3 mm × 2 mm × 0.5 mm were obtained. The morphology and elemental analyses were characterized using a scanning electron microscope (Zeiss EVO SEM) equipped with an electron microprobe analyzer for semi-quantitative elemental analysis in energy-dispersive spectroscopy (EDS) mode. Five different locations were measured using EDS. X-ray diffraction data on a single crystal at room temperature were obtained using a PANalytical Empyrean diffractometer (Cu $K_α$ radiation, $λ$ = 1.54178 Å) operated at 45 kV voltage and 40 mA current, with a graphite monochromator in a reflection mode ($2θ$ = 5°–100°, step size = 0.026°). X-ray diffraction data on ground powder from 10 K to 300 K were obtained using a HUBER imaging plate Guiner camera 670 (Cu radiation, $λ$ = 1.54059 Å) operated at 40 kV voltage and 40 mA current, with a graphite monochromator in a reflection mode ($2θ$ = 5°–100°, step size = 0.005°). Indexing and Rietveld refinement were performed using the DICVOL91 and FULLPROF programs [54]. The crystal orientation was determined using the Laue back-scattering method.

**Physical Property Measurements for Free-Strain Sample**. Temperature-dependent magnetic susceptibility and the field-dependent magnetization curves were measured using a vibrating sample magnetometer system (Quantum Design, Dynacool-14 T). Field warming curves at zero field (FW@0T) after field cooling at ±0.1 T, ±1 T, ±7 T, and field warming curves at 7 T (FW@7T) after field cooling at zero field (ZFC) and ±7 T (FC) were measured to explore out-of-plane spontaneous ferromagnetism. Then, thin flakes of $CoNb_3S_6$ with thickness from 20 $μ$m to 50 $μ$m were cleaved from the single crystal. Resistance, magneto-resistance and Hall resistance were measured using the standard six-probe configuration with the applied current (about 3 mA) along the $a$-axis. Similarly, the field warming Hall resistivity at zero field (FW@0T) after field cooling at ±1 T, ±7 T, ±14 T, and field warming curves at 7 T (FW@7T) after field cooling at ±7 T were measured. Both field-dependent magnetization and Hall resistivity measured at 30 K were taken as background and subtracted to show the evolution of out-of-plane ferromagnetic moment and anomalous Hall signals, respectively. Heat capacity measurement was carried out on the remaining crystal as temperature ranging from 2.2 K to 200 K at high vacuum (0.01 $\mu$bar).

**Hall Resistivity Measurement for Strained Sample.** Thin flakes of $CoNb_3S_6$ were mounted on a piezo-based strain cell compatible with cryostat measurements. Strain was applied both parallel and perpendicular to the *a*-axis. By changing the voltage applied to the piezo-actuator, the strain could be tuned and a standard strain gauge (gauge factor = 2.014 ± 0.002) was mounted simultaneously to calibrate the actual strain. For each strain, the Hall loops at each temperature up to 14 T were measured by cooling down from 40 K to 20 K. To test the stability of zero-field anomalous Hall resistivity, the Hall loops were also measured after saturating at different magnetic fields.

## Data Availability

The data that support the findings of this study are available from the corresponding author upon reasonable request.

## Author Contributions

J. L. and H.D.Z. conceived and oversaw the project. L.C. and H.D.Z. grew the crystals. L.C. performed the structure and physical properties characterization with the help of D.C. and A.B.. L.C. carried out the strain experiments with the help of R.L. and S.P.. All authors contributed extensively to the interpretation of the data and the writing of the manuscript.

## Competing Interests

The authors declare no competing interests.

## Acknowledgements

The authors would like to thank Prof. Y. Zhang and Prof. Y.S. Wang at Department of Physics and Astronomy, University of Tennessee, Prof. J.H. Chu at Department of Physics, University of Washington for helpful discussions. This work was supported by the Air Force Office of Scientific Research under grant no. FA9550-23-1-0502.

## References

[1] N. Nagaosa, J. Sinova, S. Onoda, A. H. MacDonald, and N. P. Ong, Reviews of Modern Physics **82**, 1539 (2010).
[2] A. Kundt, Annalen der Physik **285**, 257 (1893).
[3] W. L. Webster, Mathematical Proceedings of the Cambridge Philosophical Society **23**, 800 (1927).
[4] E. M. Pugh, Physical Review **36**, 1503 (1930).
[5] E. M. Pugh and T. W. Lippert, Physical Review **42**, 709 (1932).
[6] S. Nakatsuji, N. Kiyohara, and T. Higo, Nature **527**, 212 (2015).
[7] N. Kiyohara, T. Tomita, and S. Nakatsuji, Physical Review Applied **5**, 064009 (2016).
[8] A. K. Nayak, J. E. Fischer, Y. Sun *et al.*, Science Advances **2**, e1501870 (2016).
[9] C. Sürgers, G. Fischer, P. Winkel, and H. v. Löhneysen, Nature Communications **5**, 3400 (2014).
[10] C. Sürgers, W. Kittler, T. Wolf, and H. v. Löhneysen, AIP Advances **6** (2016).
[11] C. Sürgers, T. Wolf, P. Adelmann, W. Kittler, G. Fischer, and H. v. Löhneysen, Scientific Reports **7**, 42982 (2017).
[12] Z. Feng, X. Zhou, L. Šmejkal *et al.*, Nature Electronics **5**, 735 (2022).
[13] R. D. Gonzalez Betancourt, J. Zubáč, R. Gonzalez-Hernandez *et al.*, Physical Review Letters **130**, 036702 (2023).
[14] S.-J. Kim, J. Zhu, M. M. Piva *et al.*, Advanced Science **11**, 2307306 (2024).
[15] D. J. Thouless, M. Kohmoto, M. P. Nightingale, and M. den Nijs, Physical Review Letters **49**, 405 (1982).
[16] F. D. M. Haldane, Physical Review Letters **93**, 206602 (2004).
[17] T. Jungwirth, Q. Niu, and A. H. MacDonald, Physical Review Letters **88**, 207208 (2002).
[18] T. Jungwirth, X. Marti, P. Wadley, and J. Wunderlich, Nature Nanotechnology **11**, 231 (2016).
[19] V. Baltz, A. Manchon, M. Tsoi, T. Moriyama, T. Ono, and Y. Tserkovnyak, Reviews of Modern Physics **90**, 015005 (2018).
[20] O. Gomonay, V. Baltz, A. Brataas, and Y. Tserkovnyak, Nature Physics **14**, 213 (2018).
[21] R. Lebrun, A. Ross, S. A. Bender, A. Qaiumzadeh, L. Baldrati, J. Cramer, A. Brataas, R. A. Duine, and M. Kläui, Nature **561**, 222 (2018).
[22] K. Anzenhofer, J. M. Van Den Berg, P. Cossee, and J. N. Helle, Journal of Physics and Chemistry of Solids **31**, 1057 (1970).
[23] S. S. P. Parkin, E. A. Marseglia, and P. J. Brown, Journal of Physics C: Solid State Physics **16**, 2765 (1983).
[24] S. Mangelsen, P. Zimmer, C. Näther, S. Mankovsky, S. Polesya, H. Ebert, and W. Bensch, Physical Review B **103**, 184408 (2021).
[25] X. Zhou, W. Feng, X. Yang, G.-Y. Guo, and Y. Yao, Physical Review B **104**, 024401 (2021).
[26] P. Liu, A. Zhang, J. Han, and Q. Liu, The Innovation **3**, 100343 (2022).
[27] A. Zhang, K. Deng, J. Sheng *et al.*, Chinese Physics Letters **40**, 126101 (2023).
[28] N. D. Khanh, S. Minami, M. Hirschmann *et al.*, arXiv preprint arXiv:2403.01113 (2024).
[29] H. Tanaka, S. Okazaki, K. Kuroda *et al.*, Physical Review B **105**, L121102 (2022).
[30] N. J. Ghimire, A. S. Botana, J. S. Jiang, J. Zhang, Y. S. Chen, and J. F. Mitchell, Nature Communications **9**, 3280 (2018).
[31] G. Tenasini, E. Martino, N. Ubrig *et al.*, Physical Review Research **2**, 023051 (2020).
[32] L. Šmejkal, R. González-Hernández, T. Jungwirth, and J. Sinova, Science Advances **6**, eaaz8809 (2020).
[33] L. Šmejkal, J. Sinova, and T. Jungwirth, Physical Review X **12**, 040501 (2022).
[34] I. Mazin and P. R. X. E. The, Physical Review X **12**, 040002 (2022).
[35] L. Vistoli, W. Wang, A. Sander *et al.*, Nature Physics **15**, 67 (2019).
[36] Z. H. Zhu, J. Strempfer, R. R. Rao *et al.*, Physical Review Letters **122**, 017202 (2019).
[37] L. Šmejkal, J. Sinova, and T. Jungwirth, Physical Review X **12**, 031042 (2022).
[38] H. Takagi, R. Takagi, S. Minami *et al.*, Nature Physics **19**, 961 (2023).
[39] K. Lu, A. Murzabekova, S. Shim *et al.*, arXiv preprint arXiv:2212.14762 (2022).
[40] H. Ueda, H. Mitamura, T. Goto, and Y. Ueda, Physical Review B **73**, 094415 (2006).
[41] Y. Shirata, H. Tanaka, A. Matsuo, and K. Kindo, Physical Review Letters **108**, 057205 (2012).
[42] J.-H. Chu, H.-H. Kuo, J. G. Analytis, and I. R. Fisher, Science **337**, 710 (2012).
[43] M. C. Shapiro, A. T. Hristov, J. C. Palmstrom, J.-H. Chu, and I. R. Fisher, Review of Scientific Instruments **87** (2016).
[44] K. Siemensmeyer, E. Wulf, H. J. Mikeska, K. Flachbart, S. Gabáni, S. Mat'aš, P. Priputen, A. Efdokimova, and N. Shitsevalova, Physical Review Letters **101**, 177201 (2008).
[45] S. Yoshii, K. Ohoyama, K. Kurosawa *et al.*, Physical Review Letters **103**, 077203 (2009).
[46] Y. H. Matsuda, N. Abe, S. Takeyama *et al.*, Physical Review Letters **111**, 137204 (2013).
[47] Y. Kamiya, L. Ge, T. Hong *et al.*, Nature communications **9**, 2666 (2018).

[48] Y. Shangguan, S. Bao, Z.-Y. Dong *et al.*, Nature Physics **19**, 1883 (2023).
[49] S. Jeon, D. Wulferding, Y. Choi *et al.*, Nature Physics **20**, 435 (2024).
[50] Y. Kuramoto, H. Kusunose, and A. Kiss, Journal of the Physical Society of Japan **78**, 072001 (2009).
[51] A. Cano, M. Civelli, I. Eremin, and I. Paul, Physical Review B **82**, 020408 (2010).
[52] U. Karahasanovic and J. Schmalian, Physical Review B **93**, 064520 (2016).
[53] M. H. Christensen, J. Kang, B. M. Andersen, and R. M. Fernandes, Physical Review B **93**, 085136 (2016).
[54] J. Rodríguez-Carvajal, CEA/Saclay, France **1045**, 132 (2001).